\documentclass{JHEP3}

\usepackage{graphicx}
\usepackage{verbatim}
\usepackage{epsf}
\usepackage{latexsym}
\usepackage{amsmath,amsfonts,amssymb,amsthm}

\def\bseq{\begin{subequation}}  
\def\eseq{\end{subequation}}
\def\bsea{\begin{subeqnarray}}  
\def\esea{\end{subeqnarray}}

\newcommand{\bbox}{\lower.2ex\hbox{$\Box$}}

\newcommand{\beq}{\begin{equation}}
\newcommand{\eeq}{\end{equation}}
\newcommand{\bea}{\begin{eqnarray}}
\newcommand{\eea}{\end{eqnarray}}
\newcommand{\ena}{\end{eqnarray}}

\newcommand {\non}{\nonumber}

\newcommand{\be}{\begin{equation}}
\newcommand{\ee}{\end{equation}}

\preprint{}

\title{\begin{center} 
Pseudomoduli Dark Matter \\
and Quiver Gauge Theories
\end{center}}
\author{Antonio Amariti$^{1,a}$, Luciano Girardello$^{1,b}$,
Alberto Mariotti$^{2,c,}$\footnote{Postdoctoral researcher of FWO-Vlaanderen.} 
\\ ~
\\
$^1$Dipartimento di Fisica, Universit\`a di Milano Bicocca\\
and \\
INFN, Sezione di Milano-Bicocca,\\ 
piazza della Scienza 3, I-20126 Milano, Italy\\
\\
$^2$
Theoretische Natuurkunde, Vrije Universiteit Brussel \\
and\\
The International Solvay Institutes\\ 
Pleinlaan 2, B-1050 Brussels, Belgium\\
 ~~\\
  $^a$\email{antonio.amariti@mib.infn.it} \\
$^b$\email{luciano.girardello@mib.infn.it} \\
$^c$\email{alberto.mariotti@vub.ac.be} 
}

\abstract{ We investigate supersymmetric models for dark matter which
  is represented by pseudomoduli in weakly coupled hidden sectors.  We
  propose a scheme to add a dark matter sector to quiver gauge
  theories with metastable supersymmetry breaking.  We discuss the
  embedding of such scheme in string theory and we describe the dark
  matter sector in terms of D7 flavour branes.  We explore the phenomenology in
  various regions of the parameters.}

\begin{document}

\section{Introduction}

Cosmological observations have established the existence of dark
matter which is not composed by any of the Standard Model (SM)
particles, and with a relic abundance of the order of 
$\Omega h^2 \sim 0.1$ (for reviews see 
\cite{Jungman:1995df,Trodden:2004st,Rubakov:2005tx,Hooper:2009zm}
and reference therein).

A stable particle in thermal equilibrium with the SM in the early universe
is usually referred as cold dark matter (DM).
As long as the universe expands, the DM ceases to annihilate efficiently
and freezes out, leaving a relic abundance \cite{relicosi}
\be
\Omega h^2 \sim \frac{1}{\langle  \sigma v \rangle} 
\sim \frac{\alpha_{DM}}{M_{DM}^2}
\ee
where $h$ is the Hubble parameter in units of 100 km/sec per Mpc,
$\langle \sigma v \rangle$ is the annihilation
cross section, $\alpha_{DM}$ is the characteristic coupling
and $M_{DM}$ is the mass of the dark matter particle.
For a massive particle with weak interaction ($g_{DM} \sim 1$), i.e. a WIMP, 
the required relic abundance is 
obtained for $M_{DM} \sim 1$ TeV.
The TeV scale that naturally appears 
suggests
some relations between DM and
electroweak symmetry
breaking (EWSB), and has been dubbed
as the WIMP miracle.

A standard explanation for the origin of the electroweak 
 scale is supersymmetry breaking
 (for a phenomenological introduction to supersymmetry
 and references see \cite{Martin:1997ns}). The idea that 
supersymmetry and its breaking
should account also for the origin of the DM has prompted
an extensive investigation.
Much effort has been focused
 on the mechanism of gravity
mediation, where the lightest neutralino of the MSSM should be a
stable, viable DM candidate. 
A phenomenological drawback for models 
of gravity mediation is that they do not address
the susy flavour problem.

The flavour problem is solved in the gauge mediation scenario 
\cite{Dine:1982zb,Giudice:1998bp},
because the SM gauge interactions are flavour blind.
In gauge mediated models supersymmetry is broken at low energies
and the gravitino is the LSP.
However, the gravitino is typically 
too light (with mass in the range
of eV to Gev) \cite{Fayet:1977vd,Pagels:1981ke} 
to be a viable cold DM candidate.
Other alternatives for susy DM have
been and are under inspection (see for instance 
\cite{sviluppi,Pospelov:2007mp,Chun:2008by}).

In gauge mediated theories 
particles in the hidden sectors can
be alternative DM candidates.
In such models, 
supersymmetry breaking 
in the hidden sector occurs dynamically,
via strong dynamics effects.
The DM can then be identified with the mesons and the 
baryons of
the strongly coupled hidden sector \cite{Dimopoulos:1996gy}.

On the other hand, one can explore the possibility
of realizing dark matter in
weakly coupled hidden sector 
with spontaneous supersymmetry breaking \cite{Shih:2009he,KerenZur:2009cv}.
These models are relevant since they
can arise as the 
low energy 
description of UV free gauge theories,
as shown by \cite{Intriligator:2006dd}. 

Such weakly coupled models
typically break supersymmetry in metastable vacua,
with a spectrum of elementary particles.
The scalar potential often presents tree level flat directions,
which are lifted by one loop quantum corrections.
These pseudomoduli fields have weak scale
interactions and their one loop mass is of the order
of the supersymmetry breaking scale, i.e 
the TeV scale naturally arises.
If they are stable against decay because of 
a discrete $Z_2$ symmetry,
they represent viable cold dark matter candidates.
This possibility has been explored in O'Raifeartaigh like models in
\cite{Shih:2009he} and also analyzed in \cite{KerenZur:2009cv}.

In this paper we shall study models of pseudomoduli as dark matter and
wish to answer some questions concerning the relative UV completion.
This is a non trivial problem, with severe constraints
\cite{Adams:2006sv,Antoniadis:2007xc} for consistent embedding in a UV
complete theory.  We shall propose a quiver model for which a stringy
origin can be reached.  We embed models with pseudomoduli DM in
appropriate quiver gauge theories which arise from $D$-branes at CY
singularities.  The starting point is a quiver theory, inherited
and/or interpreted as an IR Seiberg dual, which provides the hidden
sector of dynamical breaking of susy.  The next step is the addition
of a dark matter sector, represented by an extra node in the quiver
connected through matter interactions to the hidden sector.  The
addition of $D7$ flavour branes in the CY can be put in correspondence
with the dark matter sector in the quiver.

The structure of the paper is the following. In section
\ref{pseudomoduliDM} we propose the general strategy to embed
pseudomoduli dark matter in quiver gauge theories, we comment on the
phenomenological constraints and also on the string theory
realization.  In section \ref{KOOsec} we build a concrete example; we
couple pseudomoduli DM to the KOO model \cite{Kitano:2006xg}, and we
discuss the related phenomenology.  In section \ref{stringemb} we provide for
a UV origin to the model in terms of a step of Seiberg duality: the
UV model is obtained by deforming an $L^{131}$ non isolated
singularity and by adding flavour D7 branes.  A conclusion follows.
In appendix \ref{decay} we review a basic cosmological bound on the
supersymmetry breaking scale.  In appendix \ref{albeapp} we review the
procedure of flavoring with D7 branes and we discuss its relation with
the DM sector.  In appendix \ref{APP} we provide the details of the
one loop computations
for the pseudomoduli masses.\\
~\\
~\\
As we were finishing this paper, we were informed of \cite{Inaki} which study
leptophilic dark matter \cite{Fox:2008kb} in quiver gauge theories.

\section{Pseudomoduli DM in quiver gauge theories}
\label{pseudomoduliDM}

\begin{figure}
\begin{center}
\includegraphics[width=8cm]{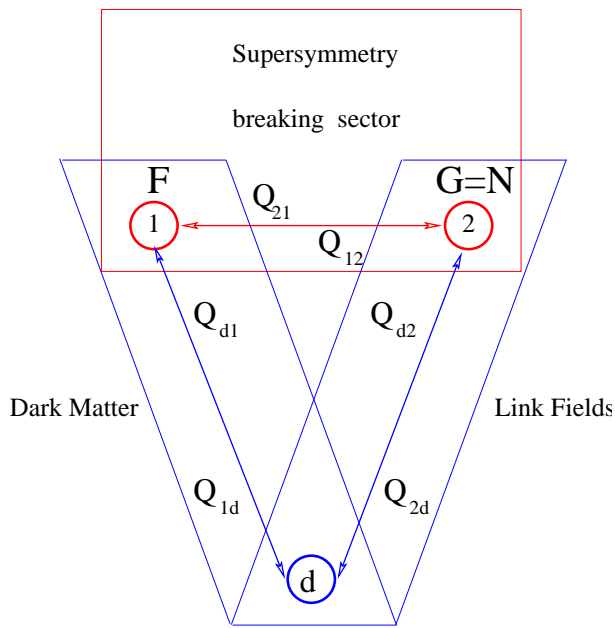}
\caption{Quiver representing the dark matter 
coupling to the supersymmetry breaking sector}
\label{generalDM}
\end{center}
\end{figure}
As anticipated, quiver gauge
theories 
are useful settings for the study of the stringy origin of
pseudomoduli dark matter.
In the quiver, we distinguish between a supersymmetry breaking sector
and a DM sector.  The first one is thought as an ISS like model: it is
characterized by a metastable vacuum, a weakly coupled $SU(N)$ gauge
symmetry and an $SU(F)$ flavour symmetry.  We also require
$R$-symmetry to be broken in this sector.
We parametrize this sector with a spurionic chiral field $X$, 
an $SU(N)$ singlet,
which acquires a vev and an $F$-component, i.e. $X=M +\theta^2 F$,
and its fermionic component is the Goldstino.
The chiral field $X$
couples to two chiral fields $Q_{12}$ and
$Q_{21}$, respectively in the fundamental and antifundamental
representation of the gauge group $SU(N)$, and in the antifundamental
and fundamental of the flavour symmetry $SU(F)$.  A three-linear coupling
$X Q_{12} Q_{21}$ induces a mass splitting between the bosonic and fermionic
components of the fields $Q_{12}$ and $Q_{21}$, which are the messengers
of susy breaking.  This sector is depicted in the rectangular region
in the figure \ref{generalDM}.

We then add the DM sector, characterized by a
$U(1)_{d}$ gauge symmetry and two pairs of bifundamental fields (see figure
\ref{generalDM}).  These fields interact with the messengers through
the superpotential
\be
\label{refpot}
W_{DM}=  Q_{12} Q_{2d} Q_{d1} +  Q_{21}Q_{1d}Q_{d2}+ m_{d2}  Q_{2d}Q_{d2}
\ee
where the traces are understood.
This new sector does not change the 
solution of the equations of motion for the hidden sector,
and the metastable vacuum remains a tree level
minimum of the scalar potential.

In the DM sector, 
the fields $Q_{2d}$ and $Q_{d2}$ have a tree level mass and
 are stabilized at zero vev. They are called link fields.

The other chiral 
multiplets $Q_{1d}$ and $Q_{d1}$ are massless at tree level.
In particular, their scalar components 
are tree level flat directions not associated with
any broken global symmetry, i.e. pseudomoduli.
Both the fermionic 
and bosonic components 
of these 
multiplets $Q_{1d}$ and $Q_{d1}$
can get a mass at one loop.
Since supersymmetry is broken,
we expect 
scalars and fermions to get different masses.
Typically, as discussed in
\cite{Shih:2009he,KerenZur:2009cv}, 
the scalar masses are higher than
the fermionic masses.
This implies that in our setting the fermions $\psi_{Q_{1d}}$ and
$\psi_{Q_{d1}}$ have lower masses than their scalar partners.
These fermions are 
the cold DM candidates.

\subsubsection*{Brief and comments}

The previous scheme can generate viable dark matter candidates
provided the hidden sector fulfills some basic requirements.

The mechanism that we consider 
for communicating the supersymmetry
breaking to the visible sector (MSSM) is gauge mediation
\cite{Dine:1982zb,Giudice:1998bp}.  This is a standard technique in
susy breaking quiver gauge theories
\cite{Kawano:2007ru,Amariti:2007am,Aharony:2007db}.
Here we shall concentrate on the direct gauge mediation scheme.

Direct gauge mediation is realized by embedding the MSSM gauge group in some
subgroup of the hidden sector.  Different choices lead to charged or
to uncharged dark matter under the MSSM gauge group.  
Dark
matter is charged if we identify the MSSM gauge group with $SU(F)$,
whereas it is uncharged if we embed the MSSM gauge
group in $SU(N)$.
Alternatively, we can give mass to the other pair of chiral fields
adding $m_{1d} Q_{1d} Q_{d1}$ to the superpotential (\ref{refpot}), setting
$m_{2d}=0$.  This exchanges the role of dark matter and link fields as
well as that one of charged and uncharged dark matter.
In this paper we discuss the case of uncharged dark matter,
which is less constrained by  experiments.
In this case the massive link fields
are charged under the MSSM gauge group
and the $U(1)_d$ dark matter gauge group.
The dark matter is charged under the $U(1)_{d}$ 
gauge group and
a hidden sector flavour group.
The interaction between the MSSM gauge group
and the dark matter is obtained via kinetic mixing 
\cite{Holdom:1985ag}.
Recently this mechanism has been investigated in 
\cite{Pospelov:2007mp,Chun:2008by}.

In models of gauge mediation $R$-symmetry has to be broken
for the gaugino to acquire non trivial mass.
The models analyzed in \cite{Shih:2009he} were required to respect a
spontaneously broken
$R$-symmetry that prevents tree level mass terms and extra couplings
for the pseudomoduli dark matter candidates.
We also
admit the explicit breaking of $R$-symmetry in the hidden
sector.  Indeed in our case the UV stringy origin uniquely settles the
structure of the interaction.

There should be a discrete $Z_2$ symmetry that forbids dark matter to
decay.  In our case this symmetry follows automatically from the
non-chiral structure of the DM sector and from its interaction
superpotential (\ref{refpot}). 
Furthermore the
pseudomoduli dark matter $Q_{2d}$ and $Q_{d2}$ have to be stabilized at one
loop at the origin of the pseudomoduli space such that this $Z_2$
symmetry is unbroken.


As discussed in \cite{Shih:2009he,KerenZur:2009cv}, we expect
the one loop masses of the DM fermions to be 
smaller 
than the masses of their scalar superpartners. This high difference between scalar
and fermion masses implies that the decay of the $Q_{2d}$ and $Q_{d2}$
scalars does not affect the dark matter relic density.  This property
of the one loop masses for the scalars and the fermions 
has to be checked in any model of pseudomoduli DM.

We require 
to have a TeV scale dark matter mass, and also
a not too heavy superpartner spectrum in th MSSM.
Hence we demand the parameter 
\be
\label{parametroR}
R \equiv \frac{M_{DM}}{m_{\lambda}} 
\ee
to be of order 1, where 
we estimate the soft mass scale
with respect to the gaugino mass $m_{\lambda}$.
The dominant contribution to the
dark matter 
cross section comes from its annihilation into dark photons
\be
\langle \sigma v \rangle \simeq \frac{\pi \alpha_{d} }{M_{DM}^2}
\ee
where $\alpha_d$ is the coupling of the 
$U(1)_d$ gauge group and $M_{DM}$ is the
dark matter mass.
For $g_d \sim 1$ and $M_{DM} \sim O$(TeV)
this cross section
secures a satisfactory relic abundance, 
and it avoids dark matter overabundance \cite{Shih:2009he,KerenZur:2009cv}.
Note that for this annihilation to be efficient
the dark gauge boson mass needs to be lower than
the dark matter mass \cite{ArkaniHamed:2008qn,Chun:2008by}. 
We will not address 
the naturalness of this new scale
here.

The realization we provided of pseudomoduli DM in gauge theories is
rather generic.  Many models of supersymmetry breaking in metastable
vacua with explicitly broken $R$-symmetry have been realized
\cite{Amariti:2006vk, Csaki:2006wi,Abel:2007jx,Intriligator:2007py,
Shih:2007av,Giveon:2007ef,Haba:2007rj,
  Giveon:2008ne,Essig:2008kz}.  
The procedure for adding a DM sector just explained
can be applied to all these models.

\subsubsection*{The stringy origin of the DM sector}
Metastable supersymmetry breaking is common to gauge theories arising
from $D3$-branes at CY singularities.  In this framework the DM sector
corresponds to $D7$-flavour branes. 
 Indeed, by properly adding
$D7$-branes, we obtain the field content and the interaction
superpotential of the DM sector. 
In the appendix \ref{albeapp} we
discuss the realization of DM sector as $D7$-branes.
This suggests a connection between pseudomoduli
DM and $D7$-branes at CY singularities.

\section{Coupling DM to the KOO model}
\label{KOOsec}
In this section we give a concrete example of the 
strategy presented above. 
We consider as the supersymmetry breaking sector the
KOO model \cite{Kitano:2006xg}. 
The low energy description of this theory
is a three node quiver gauge theory
$U(N_1) \times U(N_2) \times U(N_3)$,
where
\be
N_1=N_2=N \qquad N_3=N+M
\ee 
We choose $N<M$ such that the $U(N_2)$ gauge group is infrared free. 
The other gauge groups are considered as
very weakly coupled at low energy.
The superpotential for this model is
\be
W_{\text{KOO}} = 
-h \mu_1^2 X_{11} -h \mu_3^2 X_{33} + h m_{13} X_{11}X_{33}
+h q\cdot X \cdot \tilde q
\ee
where 
\be
q=\left(
\begin{array}{cc}
q_{21}&q_{32}
\end{array}
\right)
\quad
\tilde q=
\left(
\begin{array}{c}
q_{12}\\
q_{23}
\end{array}
\right)
\quad
X=
\left(
\begin{array}{cc}
X_{11}&X_{31}\\
X_{31}&X_{33}
\end{array}
\right)
\ee
This model breaks supersymmetry in a long living 
metastable vacuum and
does not have an $R$-symmetry.
The absence of $R$-symmetry implies that 
direct gauge mediation is a viable mechanism
to transmit supersymmetry breaking to the
superpartners in the MSSM \cite{Kitano:2006xg}.

\begin{figure}
\begin{center}
\includegraphics[width=5cm]{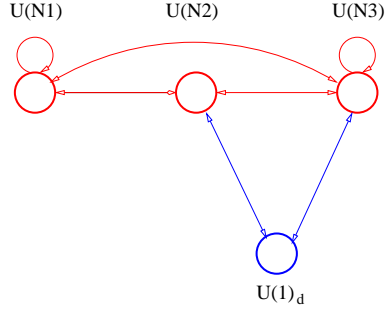}
\caption{Quiver representing the flavored KOO model}
\label{fig}
\end{center}
\end{figure}

We now add the dark matter sector. As already explained we add a new
gauge group $U(1)_{d}$ with bifundamental fields connected with the
groups $U(N_2)$ and $U(N_3)$.  The resulting quiver is depicted in
figure \ref{fig}.  
The new fields interact with the supersymmetry
breaking sector fields via the superpotential
\be
\label{defo}
W_{DM}= h q_{23}q_{3d}q_{d2}+ h q_{32}q_{2d}q_{d3}
+h m_{2} \, q_{2d}q_{d2}
\ee
The metastable supersymmetry breaking vacuum of the 
KOO model
is not destabilized by the deformation (\ref{defo}).
Hence the vacuum expectation values of the fields are
\bea
&&
q=\left(
\begin{array}{cc}
\mu_1 \mathbf{1}_{N} ~~
& 0
\end{array}
\right)
\quad
\tilde q= 
\left(
\begin{array}{c}
\mu_1 \mathbf{1}_{N}
\\
0
\end{array}
\right)
\quad
X=
\left(
\begin{array}{cc}
0&0\\
0&\chi
\end{array}
\right) 
\\
&&
q_{3d}=Y \quad 
q_{d3}=
\tilde Y
\quad
q_{2d}=0
\quad
q_{d2}=0
\eea
The vev of the fields $q_{12}$ and $q_{21}$
break the groups $U(N_1) \times U(N_2)$
to the diagonal subgroup $U(N)_{1-2}$.
The fields $\chi, Y, \tilde Y$ are pseudomoduli.
Their scalar components 
get masses via one loop corrections.
The Goldstino is a mixture of the fermionic component of $\chi$
and of $X_{11}$ \cite{Essig:2008kz,Zur:2008zg},
and
it is eaten by the gravitino in
the super Higgs mechanism \cite{Cremmer:1982en}.
The gravitino is typically too light to be 
a viable dark matter candidate in 
this gauge mediation scenario. 
The fermionic components of the fields $Y, \tilde Y$
are then natural dark matter candidates. 
We shall perform a quantitative analysis
of the spectrum to justify this scenario.

As a standard procedure we expand the superpotential around the
vacuum 
\bea
&&
q=\left(
\begin{array}{cc}
\mu_1 \mathbf{1}_{N}+ \sigma_1 ~~
& \Phi_1
\end{array}
\right)
\quad
\tilde q= 
\left(
\begin{array}{c}
\mu_1 \mathbf{1}_{N}+\sigma_2
\\
\Phi_2
\end{array}
\right)
\quad
X=
\left(
\begin{array}{cc}
\sigma_3&\Phi_3\\
\Phi_4&\chi
\end{array}
\right) 
\\
&&
q_{3d}=Y \quad 
q_{d3}=
\tilde Y
\quad
q_{2d}=\Phi_5
\quad
q_{d2}=\Phi_6
\eea
and then study the quantum infrared model which is a generalized 
O'Raifeartaigh model 
\bea 
W&=&h \chi \Phi_1 \Phi_2-h \mu_3^2 \chi+h \mu_1 
(\Phi_1 \Phi_4+\mu_1 \Phi_2 \Phi_3)+h  m_{13} \Phi_3 \Phi_4 \non \\
&&
\label{totale}
+h  Y \Phi_1 \Phi_6+ h \tilde Y \Phi_2 \Phi_5 +
h m_{2} \Phi_5 \Phi_6
\eea
The pseudomodulus $\chi$ get one loop corrections 
only from the first line in (\ref{totale}),
which is the same microscopic 
superpotential as in \cite{Kitano:2006xg}.
The second line in (\ref{totale}) is the dark matter sector.
The massive fields $\Phi_5$ and $\Phi_6$ are the link fields.
The scalars and the fermions of the fields $Y$ and $\tilde Y$
get both one loop masses.
In the appendix \ref{APP} we give the analytical
calculations of these masses.
Here we discuss the phenomenology of the model.

\subsubsection*{Phenomenology}
The stability 
of the metastable supersymmetry breaking vacuum
requires
\be
\mu_1 \gg \mu_3 \qquad  \mu_1 > m_{13}
\ee
where $\mu_1$ is the messenger mass scale, $\mu_3$
is the supersymmetry breaking scale, and
$m_{13}$ is the $R$-symmetry breaking mass.
As already explained, supersymmetry breaking
is communicated to the MSSM via direct gauge
mediation.
The GUT $SU(5)$ gauge group can be embedded
both in $U(N_3)$ or in the diagonal subgroup
$U(N)_{1-2}$.
In the first case the pseudomoduli DM 
is charged under the GUT group.
We will not investigate this possibility.
We choose the second case, leading to uncharged
DM.
The gaugino mass is \cite{Kitano:2006xg}
\be
m_{\lambda}=\frac{g^2}{16 \pi^2}(N+M) \frac{h \mu_3^2 m_{13}}{\mu_1^2} 
+O \left( \frac{m_{13}^2}{\mu_1^2} \right)
\ee
The scalar masses are of the same order provided that 
$m_{13} \sim \mu_1/\sqrt{N+M}$.

We now discuss the quantum aspects of the dark matter
sector.
The scalar component of the chiral fields 
$Y, \tilde Y$ 
acquire positive squared masses
and are stabilized at the origin of the moduli space.
This preserves the $Z_2$ discrete symmetry that makes
the DM stable.
The DM, i.e. the fermionic component of the field
$Y, \tilde Y$, also acquire one loop masses as well.

A detailed computation of the 1-loop scalar and fermionic masses
for the components of the chiral fields $Y$ and $\tilde Y$,
at all order in the supersymmetry breaking scale,
is carried out in the appendix \ref{APP}.
Here we report the analytic result at the 
combined third order 
in the adimensional 
parameters
$m_{13}/\mu_1$, $m_{2}/\mu_1$ and $\chi/\mu_1$, and
at first order in the supersymmetry breaking scale $\mu_3^2$.
The fermion mass results
\be
m_{\psi_Y \psi_{\tilde Y}}=
(N+M)
\frac{h^2}{16 \pi^2} \frac{\mu_3^2 m_{2}}{\mu_1^2} (1+\frac{2}{3} \frac{m_{13}^2}{\mu_1^2}-\frac{1}{3} \frac{\chi^2}{\mu_1^2}  )
+\dots
\ee
The scalar diagonal and off diagonal masses are respectively
\be
\label{diagmasse}
m^2_{Y Y^{*}}=
m^2_{\tilde Y \tilde Y^{*}}=
(N+M)\frac{h^2}{32 \pi^2 } \frac{\mu_3^4}{\mu_1^2} (1-\frac{m_{2}^2}{\mu_1^2}-2\frac{\chi^2}{\mu_1^2})
+\dots
\ee
\be
m^2_{Y \tilde Y}=
-(N+M)
\frac{h^2}{16 \pi^2} \frac{\mu_3^2 m_{2} \chi }{ \mu_1^4}
+\dots
\ee
The eigenvalues of the scalar masses can be obtained diagonalizing
the resulting mass matrix.
The off diagonal components are subleading and hence
the main contribution to the eigenvalues comes from the
diagonal masses (\ref{diagmasse}).
In the expressions above we should 
insert the vev of $\chi$ as a function of the other parameters,
which is found by minimizing the effective potential.

\begin{figure}
\begin{center}
\includegraphics[width=8cm]{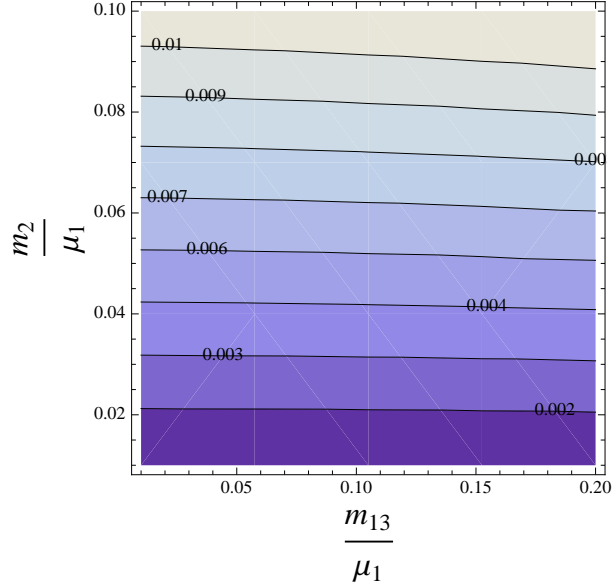}
\caption{Ratio $m_{\psi_{Y} \psi_{\tilde Y}}/m_{Y}$ where with $m_{Y}$ we denote
the lowest eigenvalue of the scalar mass matrix.
The masses are evaluated at the first order in the susy breaking scale, such that it
cancels out in the ratio. We fix the couplings to $h=1,g=1$.
}
\label{grafico1}
\end{center}
\end{figure}

Note that there should be at least one order of magnitude between
the lowest eigenvalue of the scalar mass matrix and the fermion mass,
otherwise the DM relic abundance is
affected by the decay of the scalars $Y$ and $\tilde Y$.
In the appendix \ref{decay}
we review how this constraint gives a bound on the
scale of susy breaking.
In figure \ref{grafico1} we plot the ratio of
fermion and the lowest scalar mass as a function of
the parameters of the model: it
shows that there are two order of magnitude between
the two masses. 
This constraints the susy breaking scale to 
$\mu_3 < 10^5$ TeV (see appendix \ref{decay}).

\begin{figure}
\begin{center}
\includegraphics[width=8cm]{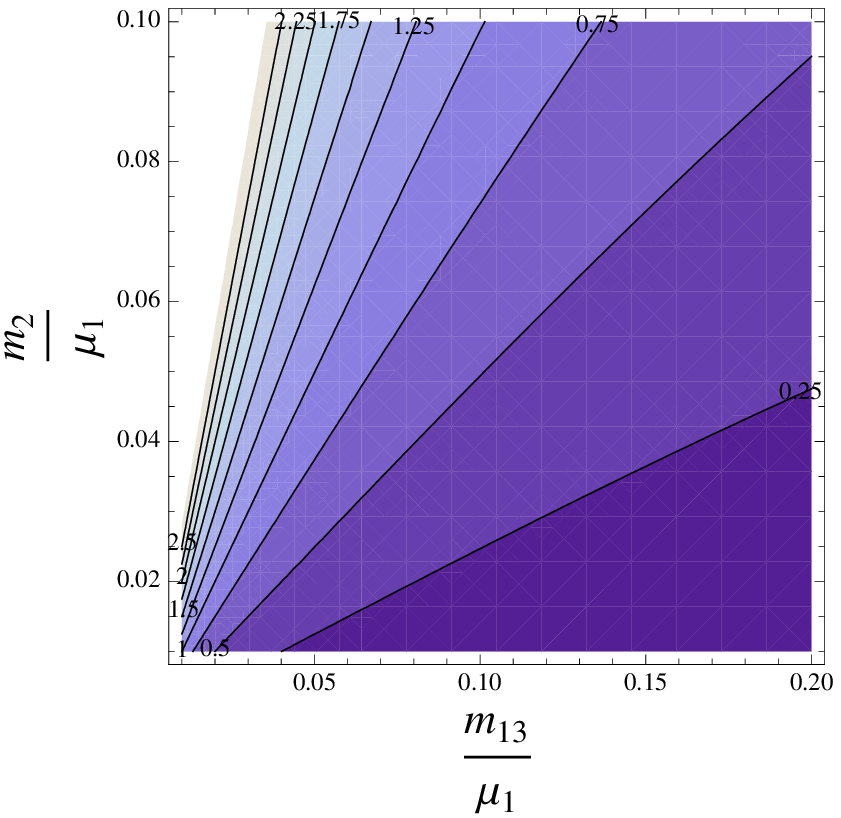}
\caption{Ratio $R$ as a function of the parameters
$m_{13}/\mu_1$ and  $m_{24}/\mu_1$.}
\label{grafico2}
\end{center}
\end{figure}

As already explained, 
we require the DM to be of the same order of the 
soft masses ($\sim 1$ TeV). 
We estimate the parameter $R$ (\ref{parametroR}) as
\be
\label{parR}
R= \frac{ m_{\psi_Y \psi_{\tilde Y}} }{m_{\lambda} }
\simeq
 \frac{h m_{2}} {g^2 m_{13}} + \dots
\ee
and we plot it in figure \ref{grafico2}.
Figure \ref{grafico2} shows that we can find a 
range in the parameter space where the ratio $R$
is of order $1$.
Notice that the enhancement factor $(N+M)$
cancels in the ratio $R$. This is a specific 
feature of the model, since the messenger fields
and the dark matter are charged under the same global symmetry
$U(N_3)$.

As usual in theories with direct gauge mediation, this
model can suffer from a Landau pole problem. 
As pointed out in \cite{Zur:2008zg},
requiring perturbative unification of the couplings
below the Landau pole forces the messenger scale and the
supersymmetry breaking scale to be large. This results in
a too large mass for the gravitino, outside the cosmological bound
worked out in \cite{Viel:2005qj}.
We leave a detailed analysis of the issue of gauge coupling unification
for future studies.
One can solve this problem by looking at a different UV completion for
the KOO model, for example via a cascading gauge theory.
In the next section we show how the same low energy theory 
arises from a system of $D3$ and $D7$ branes probing a CY singularity
through a Seiberg duality.

Alternatively, one can choose a different embedding for the SM gauge
group into the flavour group of the supersymmetry breaking sector.
This embedding of $SU(5)_{GUT}$ into $U(N_3)$ has been investigated
in \cite{Zur:2008zg}, and it has been shown to be compatible with 
a 
gravitino mass which is consistent with the cosmological bound of
\cite{Viel:2005qj}.
For this embedding the pseudomoduli $Y$ and $\tilde Y$ are charged under the
standard model gauge group. If we want to 
realize 
DM uncharged under the SM gauge group, we have to 
exchange the role of 
the dark matter and of the link fields. This is done by
setting to zero 
the mass term in
(\ref {defo}), and adding a new one $h m_{3} q_{3d}q_{d3}$.
The phenomenology of the low energy theory 
and the ratio $R$ are unchanged.

\section{Pseudomoduli DM from $D$-brane at CY singularities}
\label{stringemb}
String theory provides a 
natural embedding for pseudomoduli DM. 
In fact, deformations of non isolated singularities 
lead to metastable supersymmetry breaking
in the gauge theory living on 
$D3$-branes probing the singularities 
\cite{braneschi,Franco:2006es}.
It has also been shown \cite{Franco:2006es} that
the addition of flavour $D7$ branes 
can break supersymmetry in metastable vacua.
Here we use both these effects
to build a quiver
gauge theory.
We show that 
the resulting model
is a concrete
realization of the setup
for pseudomoduli DM 
in quiver gauge theories 
proposed in section \ref{pseudomoduliDM}.

\begin{figure}
\begin{center}
\includegraphics[width=5cm]{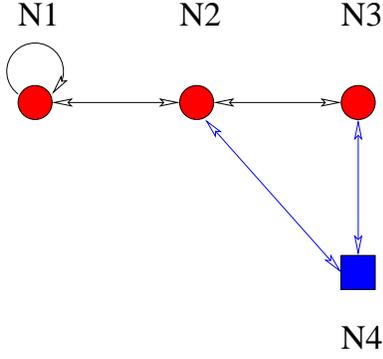}
\caption{Quiver representing the flavored $L^{131}$ model.}
\label{L131ele}
\end{center}
\end{figure}

We consider the quiver gauge theory arising 
from $D3$ branes at the $L^{131}$ singularity.
It can be described by the curve in $C^4$
\be
x y^3 = wz
\ee
and we deform it as
\be
\label{defgeom}
x (y+\xi) y (y-\xi) =w z
\ee
We choose the rank of the four gauge group as 
\be
N_0=0 \qquad N_1=1 \qquad N_2=M \qquad N_3=N
\ee
and the resulting superpotential is
\be
W_0=X_{11} Q_{21} Q_{12}+\lambda Q_{12}Q_{23}Q_{32}Q_{21}+
m_{23} Q_{23}Q_{32}
\ee
where the mass is related to the parameter in the geometry
 (\ref{defgeom}) as
$m_{23}= \xi \lambda$.
We add $N_4=M-1$ $D7$-flavour branes 
associated with the fields $Q_{23}$ and $Q_{32}$
(see appendix \ref{albeapp}). This introduces
new fields that interacts via the superpotential
\be
W_{D7}=\rho Q_{23}Q_{34}Q_{42}+\rho Q_{32}Q_{24}Q_{43}+
m_{34} Q_{34}Q_{43}+m_{24} Q_{24} Q_{42}
\ee
In figure \ref{L131ele} we give a pictorial 
quiver representation of the model,
with complete superpotential
\be
W=W_0+W_{D7}
\ee

\begin{figure}
\begin{center}
\includegraphics[width=5cm]{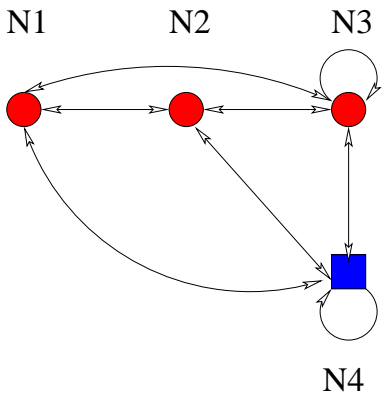}
\caption{Quiver representing the flavored $L^{131}$ model after Seiberg duality on node 2.}
\label{magn}
\end{center}
\end{figure}

We work in the range $M>2N$. In this window
the node $2$ has the strongest coupling
and it is in the magnetic free window. 
The low
energy description
can be obtained via Seiberg duality.
The dual low energy theory 
has the superpotential
\bea
W=&&Tr\left(
\begin{array}{ccc}
q_{21}&q_{23}&q_{24}
\end{array}
\right)
\left(
\begin{array}{ccc}
M_{11}&M_{13}&M_{14}\\
M_{31}&M_{33}&M_{34}\\
M_{41}&M_{43}&M_{44}
\end{array}
\right)
\left(
\begin{array}{c}
q_{12}\\
q_{32}\\
q_{42}
\end{array}
\right)
+m_{11} M_{11}X_{11}
+m_{13} M_{13}M_{31}
\nonumber \\
+&&
+m_{Q34} Q_{34}Q_{43}
+m_{MQ34} M_{34}Q_{43}
+m_{MQ34} Q_{34}M_{43}
-\mu_3^2 M_{33} 
- \mu_4^2 M_{44}
\eea
where the new scales are 
\be
m_{11}=\Lambda_2, \quad m_{13}= \lambda \Lambda_2^2, 
\quad m_{Q34}=m_{34}, \quad m_{MQ34}=\rho \Lambda_2,
\quad \mu_3^2 =m_{23} \Lambda_2,
\quad \mu_4^2 = m_{24} \Lambda_2
\ee
We work in the regime where
$m_{11}$ and $m_{Q34}$
are larger than the other scales of the theory.
We can then integrate out
the fields $X_{11}$,$M_{11}$,$Q_{34}$ and $Q_{43}$,
and obtain the superpotential
\bea
W=&&
q_{21}M_{13}q_{32}+
q_{21}M_{14}q_{42}+
q_{23}M_{31}q_{12}+
q_{24}M_{41}q_{12}+
q_{23}M_{33}q_{32}+
q_{23}M_{34}q_{42}\non\\
+&&
q_{24}M_{43}q_{32}+
q_{24}M_{44}q_{42}+
m_{13} M_{13}M_{31}+
m_{34} M_{34}M_{43}-
\mu_3^2 M_{33} - 
\mu_4^2 M_{44}
\eea
where we define $m_{34}=m_{MQ34}^2/m_{Q34}$.
Figure \ref{magn} is a pictorial 
quiver representation of the magnetic model.
The ranks of the groups in the magnetic theory are
\be
N_1=1, \quad \tilde N_2=N_3=N, \quad N_4=M-1
\ee
We look for the metastable vacuum states. 
We find
\bea
&&M_{13}=M_{31}=0,\quad
M_{34}=M_{43}^T=
0
,\quad
M_{14}=Y
,\quad
M_{41}=
\tilde Y
\\
&&
M_{33}=0
,\quad
M_{44}=\chi
,\quad
q_{12}=q_{21}=0,\quad
q_{23}=q_{32}=\mu_3,\quad
q_{24}^T=q_{42}=
0
\non
\eea
The one loop correction for the pseudomoduli are calculated after 
expanding the fields of the theory around their 
expectation value.
Nevertheless some of their fluctuations give a 
supersymmetric contribution at one loop.
The only relevant expansions are
\bea
&&q_{42}=
\phi_1
\quad
q_{24}=
\phi_2
\quad
q_{12}=\phi_{5}\quad q_{21}=\phi_{6}
\quad
M_{13}=\phi_{7}\quad M_{31}=\phi_{8}
\\&&
M_{14}=Y
\quad
M_{41}=\tilde Y
\quad
M_{34}=\phi_4
\quad
M_{43}=\phi_3
\quad
M_{44}=\chi
\non
\eea
Expanding around the vacuum we find the following 
structure for the effective superpotential
involving the pseudomoduli $\chi$, $Y$ and $\tilde Y$
\be
W_{eff}=W(\chi)+W(Y,\tilde Y)
\ee
where 
\be
W(\chi)=\chi \phi_1 \phi_2- \mu_4^2 \chi + \mu_{3}(\phi_1 \phi_4 + \phi_2 \phi_{3})
+m_{34}\phi_3\phi_{4}
\ee
and the superpotential for the other pseudomoduli is 
\be
W(Y,\tilde Y) = 
Y \phi_1 \phi_6 + \tilde Y \phi_2 \phi_5 + 
\mu_3 (\phi_6 \phi_7 + \phi_5 \phi_8)
+m_{13} \phi_7 \phi_8  
\ee
This superpotential reduces to
(\ref{totale}) in the limit $m_{13} > \mu_3$.
Indeed in this limit we can
integrate out  supersymmetrically
the fields $\phi_7$ $\phi_8$ and obtain
an effective mass term for the fields $\mu_3^2/m_{13}\, \phi_5  \phi_6$.
Hence in this limit the phenomenology of the model is the same as
in section \ref{KOOsec}.
However, also the case with $m_{13} < \mu_3$ is phenomenologically
viable, with the parameter $R$ of order 1.
These features are manifest in 
figure \ref{fighi} where we plot
the parameter $R$ as a function of the ratios $m_{13}/\mu_3$ and
$m_{34}/\mu_3$.
In appendix \ref{APP} we perform the explicit computation
for the 1-loop DM fermion mass.

\begin{figure}
\begin{center}
\includegraphics[width=10cm]{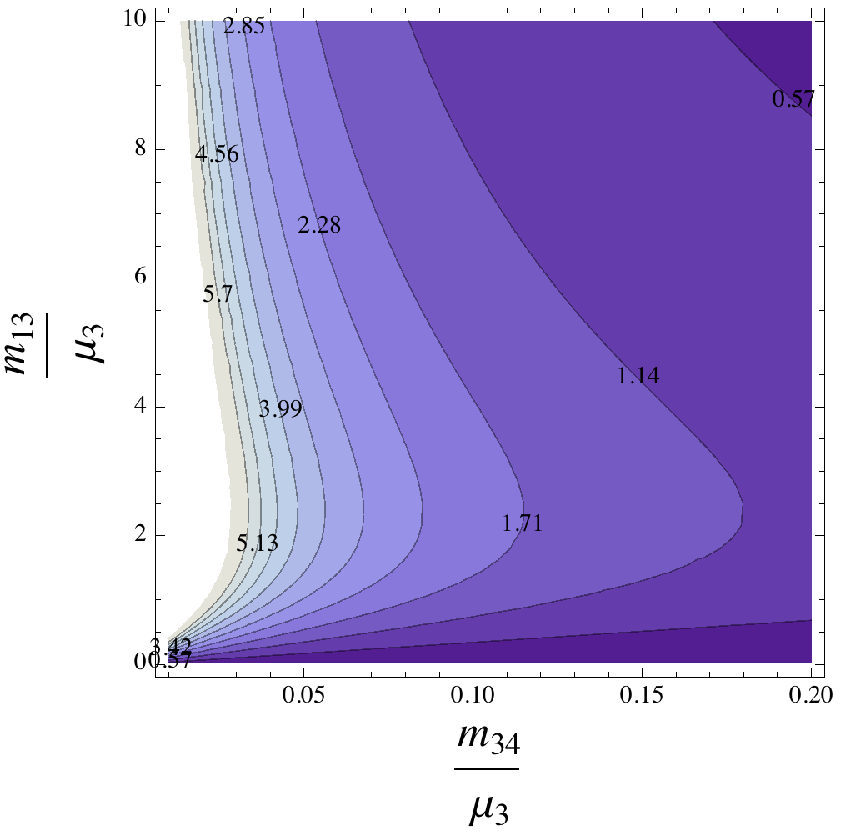}
\caption{Ratio $R$.}
\label{fighi}
\end{center}
\end{figure}

\section{Conclusions}
In this article we proposed a scheme to realize
pseudomoduli 
dark matter in quiver gauge theories
with weakly coupled supersymmetry breaking a' la ISS.
In section \ref{pseudomoduliDM} we distinguished a metastable 
supersymmetry breaking sector and a dark matter sector.
The former communicates the breaking
of supersymmetry to the MSSM trough gauge interactions. 
This mechanism requires
$R$-symmetry to be broken, spontaneously or explicitly. 
The dark matter sector
contains two pseudomoduli fields and
it is coupled to the supersymmetry breaking sector
through superpotential terms. 
This coupling induces, at one loop,
supersymmetry breaking masses for the pseudomoduli 
and their fermionic partners 
in the dark matter sector.
This provides the mechanism to generate 
a TeV scale mass for the fermions which are
the cold dark matter candidates.

We considered only the case of uncharged dark matter with
respect to the
gauge interactions of the standard model. 
There is a coupling of the $U(1)_{d}$
gauge group, under which DM is charged, to the $U(1)_Y$ of the
standard model via the kinetic mixing.
Cases in which the dark matter is charged under a non abelian gauge
group are left for future works.

We showed that when the supersymmetry breaking sector is described by
$D3$ branes probing CY singularity, the dark matter sector corresponds
to the addition of $D7$ flavour branes.

In section \ref{KOOsec}
we realized the proposed scenario by
coupling a pseudomoduli DM sector to 
the KOO model,
and $1$TeV mass dark matter is rather natural.
In section \ref{stringemb}
we studied a model with 
D$3$ and D$7$ branes wrapped over a deformed 
CY $L^{131}$ singularity.
We have argued that this model reduces, 
in the infrared, to the KOO model plus DM.

Many extensions of our proposal can be studied.
One can find other models by coupling 
the DM sector to gauge theories different from KOO.
For example one can couple DM to models with  $R$-symmetry breaking
in metastable vacua 
\cite{Amariti:2006vk, Csaki:2006wi,Abel:2007jx,Intriligator:2007py,
Shih:2007av,Giveon:2007ef,Haba:2007rj,
  Giveon:2008ne,Essig:2008kz}. 
Another interesting issue is the investigation at large of the
UV completion. The generation of the DM sector 
inside a quiver gauge theory can be a result of a step
of Seiberg duality, as we showed in the deformed and flavored
$L^{131}$ theory.

\section*{Acknowledgments}
We are grateful to Angel Uranga for 
useful comments.
We thank Vijay
Balasubramanian, Per Berglund and Inaki Garcia Etxebarria
for informing us about their research \cite{Inaki} on related topics.

A.~A.~ and L.~G.~ are supported in part by INFN,
in part  by
MIUR under contract 2007-5ATT78-002
and in part by 
the European
Commission RTN programme MRTN-CT-2004-005104.
A. ~M. ~ is
supported in part 
by the Belgian Federal Science Policy Office 
through the Interuniversity 
Attraction Pole IAP VI/11 and by 
FWO-Vlaanderen through project G.0428.06.


\appendix

\section{Cosmological bounds} \label{decay}
Pseudomoduli DM arise from a tree level massless chiral
multiplet $Y=(\phi_Y,\psi_Y)$. 
It includes a complex scalar and a fermion.
Both fields acquire masses at one loop.
In the model we studied there 
is a hierarchy among these masses. The
scalar mass is typically one or two order of
magnitude larger than the fermion mass.
There is a $Z_2$ discrete symmetry under which
$Y$ is charged. This prevents the
field $\psi_Y$ to decay.
If it has weakly coupled
interactions and mass at the TeV scale,
it is a viable DM candidate.

The scalar component, which is heavier, 
can decay in a gravitino and a fermion,
$\phi_Y \to G \psi_Y$. 
This decay can modify the relic density of
the DM candidate $\psi_Y$.
We thus require that the 
decay temperature of the scalar $\phi_Y$
is larger than the freeze out temperature of
the fermionic DM $\psi_Y$.
The scalar decay rate is 
\cite{Martin:1997ns}
\be
\Gamma (\phi_Y \to \psi_Y \, G )=\frac{m_{\psi_Y}^5}{16 \pi f^2} 
\left( 1- \frac{m_{\psi_Y}^2}{m_{\phi_Y}^2} \right)^4
\simeq 
\frac{m_{\psi_Y}^5}{16 \pi f^2} 
\ee
The temperature associated to this decay is $T \sim \Gamma^{1/2}$
\be
T_{\phi_Y}=\left( \frac{m_{\psi_Y}}{ 100 \text{TeV}} \right)^{5/2} 
\left( \frac{10^4 \text{TeV}}{\sqrt{f}}\right)^2 ~ 3 \, \text{TeV}
\ee
whereas the freeze out temperature of the fermion 
is $T_{\text{freeze}}\simeq m_{\psi_Y}/20$.
The requirement $T_{\phi_Y}>T_{freeze}$ translates
in a bound on the supersymmetry breaking scale
\be
\sqrt{f} \lesssim 
\left( \frac{m_{\phi_Y}}{100 \text{TeV}} \right)^{5/4} 
\left( \frac{1 \text{TeV}}{ m_{\psi_Y}} \right)^{1/2}
~ 2,5 \cdot 10^5 \text{TeV}
\ee

\section{Flavoring with $D7$ branes}\label{albeapp}
In this paper we have proposed how to realize a
dark matter sector in quiver gauge theories.
Here we show how this new sector can
be described as $D7$-flavour branes \cite{Karch:2002sh}.

We briefly review the technique
introduced in \cite{Franco:2006es} to add $D7$ branes to
toric quiver gauge theories and to extract the
interaction superpotential;
we refer to the original paper for 
a detailed explanation.

Consider a toric quiver gauge theory realized
as $D3$ branes probing a toric CY singularity.
The system can be described in terms of a dimer diagram
and a useful tool is the Riemann surface in the mirror 
configuration \cite{Feng:2005gw}.

By the use of these tools, it turns out that in the quiver 
we can associate to every bifundamental field 
a  supersymmetric four cycle,
which passes through the singular point, 
on which a $D7$-brane can be wrapped.
Call one of this cycle $\Sigma_{ij}$ and 
label the two gauge groups under which the bifundamental
is charged as $U(N_i)$ and $U(N_j)$.
When adding $D7$-branes, one should control
the cancellation of RR tadpoles, corresponding, 
on the field theory side, to an anomaly free theory.

\begin{figure}
\begin{center}
\includegraphics[width=6cm]{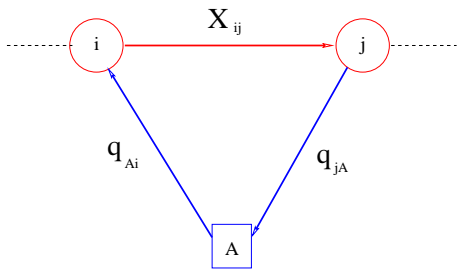}
\caption{Quiver obtained by adding a $D7$ brane associated with 
the field $X_{ij}$. This is only a sector of an anomaly free theory.}
\label{app1}
\end{center}
\end{figure}

The addition of the $D7$-brane adds new
bifundamental
fields 
corresponding to strings stretched
between the $D7$ brane and the $D3$ branes.
These new degrees of freedom are charged
under the gauge groups $U(N_i)$ or $U(N_j)$
and under 
the $U(1)_A$ symmetry introduced  
by the $D7$ brane
\footnote{If there are $K$ $D7$ branes on the same cycle and
with the same Chan-Paton structure the
symmetry is $U(K)_A$.}.
We show in the figure \ref{app1} the resulting quiver.

There is an interaction superpotential term
of the type $33-37-73$
that can be obtained analyzing the
disk on the mirror Riemann surface, and it is
\be
\label{interapp}
W_{int}= X_{ij}q_{jA}q _{Ai}
\ee
If there are $D7$ branes on different four cycles,
each of them introduces a couple of bifundamental fields
and interaction terms as in (\ref{interapp}). 
There are also interaction terms of the type $37_A-7_A 7_B -7_B 3$.
These terms can lead to masses for the $37$ fields
if a $7_A 7_B$ field get a non trivial vacuum expectation values.
The vev 
breaks the $U(1)_A \times U(1)_B$ groups
associated to the $D7_A$ and $D7_B$ branes 
to the diagonal subgroup. 

\begin{figure}
\begin{center}
\includegraphics[width=6cm]{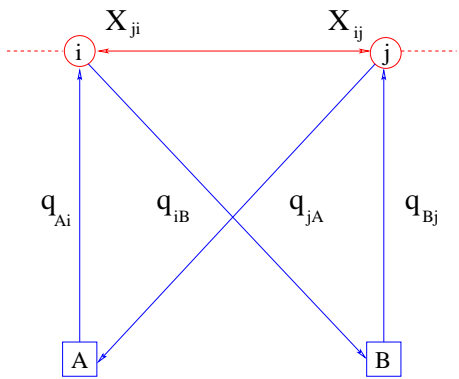}
\caption{Quiver obtained adding $D7$ brane associated to the field $X_{ij}$
and $X_{ij}$.}
\label{app2}
\end{center}
\end{figure}

For instance consider the addition of $D7$-branes associated
to two different bifundamentals $X_{ij}$ and $X_{ji}$
charged under
the group $U(N_i)$ and $U(N_j)$. 
The resulting theory is the quiver depicted in figure \ref{app2} .
A vev for the $7_A 7_B$ field give raise to the following
mass terms in the superpotential, that
involves both set of flavours,
\be
\label{massemer}
W_{mass}=m_1 q_{Ai}q_{iB}+ m_2 q_{Bj}q_{jA}
\ee
in addition to the interaction superpotential
\be
W_{int}=X_{ij}q_{jA}q _{Ai}+X_{ji} q_{iB}q_{Bj}
\ee
The mass parameters can be related to 
geometrical quantities as follows.
The mass term 
corresponds in the geometry to the recombination
of two $D7_A$ and $D7_B$ branes. The two
cycles recombine in one cycle which passes at some distance
$\epsilon$ from
the singular point. 
Taking local holomorphic coordinates we can parametrize the
two four-cycle as
$z_1=0$ and $z_2=0$. 
The mass term corresponds to a recombination
of the $D7_A$ and $D7_B$ brane such that they
now wrap the cycle $z_1 z_2=\epsilon$. The parameter
$\epsilon$ is the distance of the four-cycle from
the singular point, and it is related to
the gauge theory parameters as
\be
\epsilon \sim m_1 m_2
\ee
Then the two mass parameters have to be 
both turned on.

In the paper we used this technique to add
flavours to quiver gauge theories and to 
extract their interaction superpotential. 
In section \ref{pseudomoduliDM} 
we add $D7$-branes associated to the
fields $Q_{12}$ and  $Q_{12}$. 
We introduced only one of the two mass terms of (\ref{massemer}). 
This approximation can be considered as a 
limiting case where there is a large hierarchy among the two masses,
and hence we neglect one of the two.
This is generally the approximation to adopt 
when adding a
$D7$-brane dark matter
sector to a weakly coupled 
supersymmetry breaking sector,
as in section \ref{pseudomoduliDM} and in section \ref{KOOsec}.

In section \ref{stringemb} we analyzed a UV complete theory, and there
we introduced $D7$-flavour branes associated with the fields $Q_{23}$
and $Q_{32}$ both, with the mass terms of (\ref{massemer}). The setup
of section \ref{pseudomoduliDM} is then obtained as the low energy
description of the model by performing Seiberg duality.

\section{An explicit calculation}\label{APP}

In this appendix we show a detailed calculation of the one loop 
masses for pseudomoduli DM.
We consider the superpotential
\be \label{simplest}
W_1 = f X + X \phi_1 \phi_2 + \mu (\phi_1 \phi_3 + \phi_2 \phi_4 )
+ m_1 \phi_3 \phi_4 
\ee
This superpotential represents the supersymmetry
breaking sector. Here $R$-symmetry is explicitly broken by 
the term $m_1 \phi_3 \phi_4$.
In the non supersymmetric minimum the 
fields $\phi_i$ have zero vev. The fields $X$ 
is a pseudomodulus. 
The pseudomoduli space is tachyon free and stable if
\be
|\mu^2\pm m_1 X|^2-f (m_1^2+\mu^2)>0
\ee
The one loop analysis shows that this pseudomodulus
is stabilized at $\langle X \rangle \neq 0$ 
\cite{Kitano:2006xg,Zur:2008zg}.

\subsubsection*{DM and KOO model}

The model of section \ref{KOOsec}
is recovered by adding to (\ref{simplest})
the superpotential for the dark matter sector. This is 
\be
W_2= m_2 \phi_5 \phi_6 + Y \phi_1 \phi_5 + \tilde Y
\phi_2 \phi_6 
\ee
The fields $\phi_5$ and $\phi_6$ are stabilized at 
zero vev in the non supersymmetric vacuum, while
the fields  $Y$ and $\tilde Y$ are pseudomoduli.

They are stabilized at one loop at the origin of the moduli space, and
their scalar components and their fermions components get both a non
zero mass.
The fermions $\psi_{Y}$ and $\psi_{\tilde Y}$  get one loop masses, 
differently from $\psi_{X}$, because they are not associated with the goldstino.
However their masses
are generically different from the masses of their bosonic partners, because
they feel the effects of supersymmetry breaking. 

\begin{figure}
\begin{center}
\includegraphics[width=10cm]{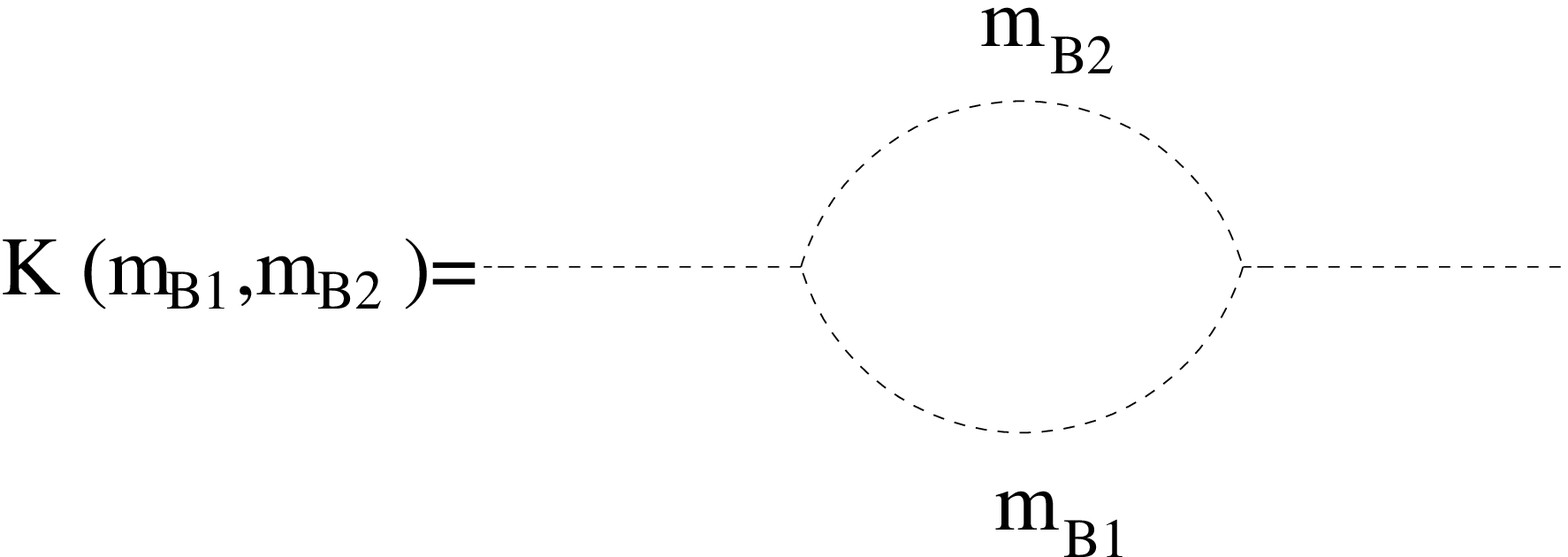}
\caption{The one loop Feynman diagram associated with the function K($m_{B1},m_{B2}$)}
\label{massascala1}
\end{center}
\end{figure}

\begin{figure}
\begin{center}
\includegraphics[width=8cm]{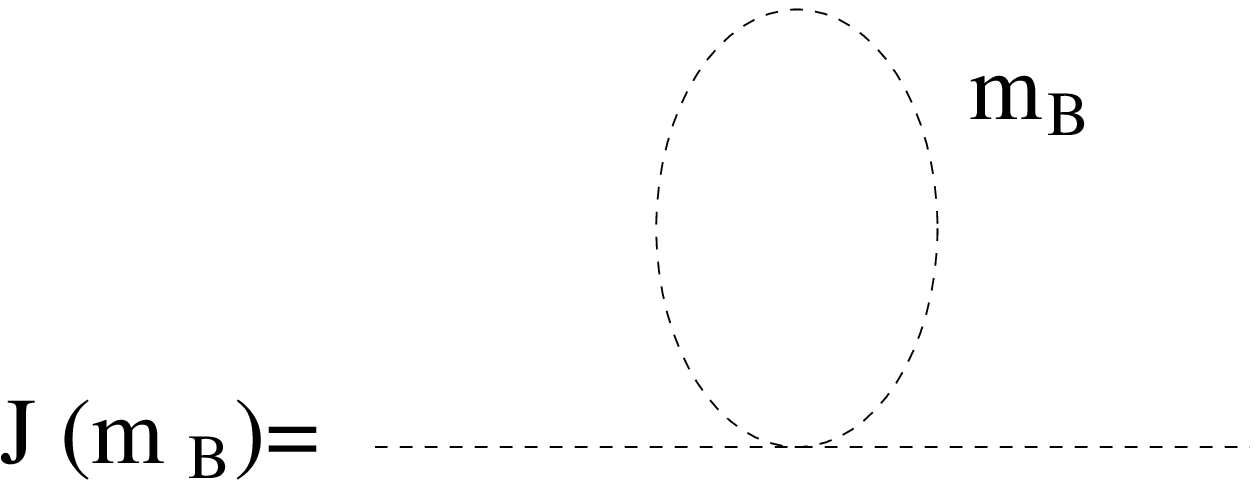}
\caption{The one loop Feynman diagram associated with the function J($m_B$)}
\label{massascala2}
\end{center}
\end{figure}

In this appendix we explicitly calculate the scalar and fermion masses.
The calculations are performed by using the approach of 
\cite{Giveon:2008wp}.
It consists of calculating the mass term for the model with $f$ turned on
and repeat the same calculation, but with $f=0$.
The two models, the one with $f\neq0$ and the one with $f = 0$,
have the same interactions, the same field content, but a different
spectrum. Since the masses of the fields get corrected only in the non
supersymmetric case, the difference between the mass in the non
supersymmetric case and the mass in the supersymmetric case coincide
with the mass in the non supersymmetric case. Trivially we can write
the equation
\be
m^{(1)}_{f\neq 0}=m^{(1)}_{f\neq 0}-m^{(1)}_{f=0}
\ee
This trick reduces the number of diagrams necessary 
to calculate the masses of the pseudomoduli.
Indeed the only diagrams that contribute 
to the mass are the ones depending on $f$ in the 
non-supersymmetric case.\\
We first calculate the fermion mass term in the effective Lagrangian 
of the form
\be
\mathcal{L}_{eff} \supset
M_{\psi} \psi_{\tilde Y} \psi_{Y}+h.c.
\ee
This term arises from the one loop diagrams due to the interactions
\be
\mathcal{L}
\supset
\psi_{Y} (\psi_5 \phi_1 + \phi_5 \psi_1) 
+ 
\tilde \psi_{Y} (\psi_6 \phi_2 + \phi_6 \psi_2) +h.c.
\ee
The calculation is not immediate, since the mass matrix
for the fields $\phi_1$ and $\phi_2$ are not diagonal. We have to 
diagonalize the bosonic mass matrices for $ X \neq 0$ and 
$Y=\tilde Y=0$.
For simplicity we first diagonalize the fermionic squared mass matrix, 
and then we diagonalize the bosonic one, breaking the holomorphic structure.
The eigenvalues of the fermionic mass matrix for the fields 
$\phi_1,\dots,\phi_4$ are
\be
m_F^{2\pm}=\frac{ m_1^2+X^2+2\mu^2\pm\sqrt{(m_1^2-|X|^2)^2+4\mu^2(m_1^2+2 \ \text{Re}(X) m_1 + |X|^2)}}{2}
\ee
The diagonal combination of the fields in the superpotential are
\bea
&&\phi_1= ~~\cos\tau \rho_1+\sin \tau \rho_4, \quad
\phi_2= ~~\cos\tau \rho_2+\sin \tau \rho_3  \non\\
&&\phi_3= -\sin\tau \rho_2+\cos \tau \rho_3,  \quad
\phi_4= -\sin\tau \rho_1+\cos \tau \rho_4
\eea  
where
\be
\sin^2\tau = \frac{\mu^2+m^2-m_F^{2-}}{m_F^{2+}-m_F^{2-}}
\ee
and
\be
m_{\psi(\rho_1)}^2=m_{\psi(\rho_2)}^2=m_F^{2-}
\ \ \ \ 
m_{\psi(\rho_3)}^2=m_{\psi(\rho_4)}^2=m_F^{2+}
\ee
Since supersymmetry is broken, the complex chiral fields
$\rho_i$ do not have holomorphic masses. It is necessary to find
the combinations of these fields that diagonalize the bosonic mass
matrix.
The eigenvalues of this matrix are
\be
m_B^{2\eta\rho}=
\frac{ 
-\eta f + m_1^2 + X^2 + 2\mu^2 +\rho\sqrt{(\eta f + m_1^2-|X|^2)^2+4\mu^2(m_1^2+|X|^2+2 \ \text{Re}(X) m_1)}
}{2}
\ee
where $\eta=\pm 1$ and $\rho=\pm 1$.
The scalar components that diagonalize the boson mass matrix are
\bea
&&
\xi_1^A=
-\text{Im}(\rho_1\!+\!\rho_2)\cos{\theta}\!+\!\text{Im}(\rho_3\!+\!\rho_4)\sin{\theta}
,\quad
~\xi_2^A=
-\text{Re}(\rho_1\!-\!\rho_2)\cos{\theta}\!-\!\text{Re}(\rho_3\!-\!\rho_4)\sin{\theta}
\non \\
&&
\xi_1^B=\phantom{-}
\text{Im}(\rho_1\!+\!\rho_2)\sin{\theta}\!+\!\text{Im}(\rho_3\!+\!\rho_4)\cos{\theta}
,\quad~\hspace{-.09mm}
\xi_2^B=\phantom{-}
\text{Re}(\rho_1\!-\!\rho_2)\sin{\theta}\!-\!\text{Re}(\rho_3\!-\!\rho_4)\cos{\theta}
\non \\
&&
\xi_1^C=\phantom{-}
\text{Im}(\rho_1\!-\!\rho_2)\cos{\gamma}\!-\!\text{Im}(\rho_3\!-\!\rho_4)\sin{\gamma}
,\quad\hspace{.57mm}
\xi_2^C=\phantom{-}
\text{Re}(\rho_1\!+\!\rho_2)\cos{\gamma}\!+\!\text{Re}(\rho_3\!+\!\rho_4)\sin{\gamma}
\non \\
&&
\xi_1^D=
-\text{Im}(\rho_1\!-\!\rho_2)\sin{\gamma}\!-\!\text{Im}(\rho_3\!-\!\rho_4)\cos{\gamma}
,\quad\hspace{.1mm}
\xi_2^D=
-\text{Re}(\rho_1\!+\!\rho_2)\sin{\gamma}\!+\!\text{Re}(\rho_3\!+\!\rho_4)\cos{\gamma}
\non \\\eea
where 
\bea
&&
\sin^2\theta=
\frac
{
f(m_1^2-X^2)
+
(m_f^{2-}-m_f^{2+})
(m_B^{2++}-m_B^{2+-}+m_f^{2-}-m_f^{2+})}
{2(m_B^{2++}-m_B^{2+-})(m_f^{2-}-m_f^{2+})}
\non \\
&&
\cos^2\gamma=
\frac
{
f(m_1^2-X^2)
+
(m_f^{2-}-m_f^{2+})
(m_B^{2-+}-m_B^{2--}-m_f^{2-}+m_f^{2+})}
{2(m_B^{2-+}-m_B^{2--})(m_f^{2-}-m_f^{2+})}\non \\
\eea
The diagonal masses of these fields are
\be
m_{A}^2=m_B^{2+-}
\ \ \
m_{B}^2=m_B^{2++}
\ \ \
m_{C}^2=m_B^{2--}
\ \ \
m_{D}^2=m_B^{2-+}
\ \ \
\ee
\begin{figure}
\begin{center}
\includegraphics[width=10cm]{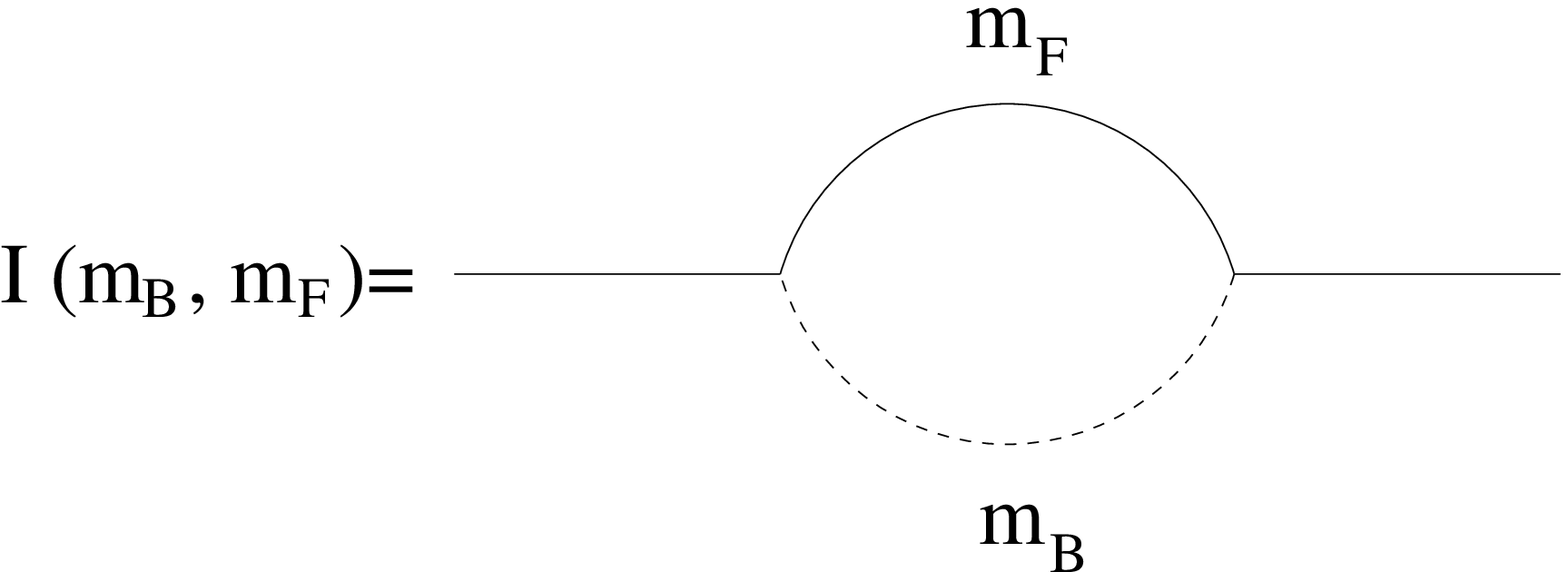}
\caption{The one loop Feynman diagram associated with the function I($m_B,m_F$)}
\label{massaferm}
\end{center}
\end{figure}
We can now evaluate the one loop fermion mass, by using the 
diagram in figure \ref{massaferm}. From this diagram one computes 
the function 
I$(m_B,m_F)$, which is
\be
I(m_B,m_F)=-\frac{m_F}{16 \pi^2}\left(
\log\frac{\Lambda^2}{m_F^2}-\frac{m_B^2}{m_B^2-m_F^2}
\log\frac{m_B^2}{m_F^2}
\right)
\ee
and the mass for the fermion is
\bea
M_{\psi_Y \psi_{\tilde Y}} =&& 
(\cos\theta \cos \tau - \sin \theta \sin \tau)^2 I(m_A,m_2)
+(\sin\theta \cos \tau + \cos \theta \sin \tau)^2 I(m_B,m_2)
\non \\ 
-&&(\cos\gamma \cos \tau + \sin \gamma \sin \tau)^2 I(m_C,m_2)
-(\sin\gamma \cos \tau - \cos \gamma \sin \tau)^2 I(m_D,m_2)
\non 
\eea
Analogously we can evaluate the mass term acquired at one loop
by the scalars $Y$ and $\tilde Y$.
The mass term in the Lagrangian is
\be
\mathcal{L} = m_{Y Y^*}^2 |Y|^2 + m_{\tilde Y \tilde Y^*}^2 |\tilde Y|^2
+(m_{Y \tilde Y} Y \tilde Y + cc)
\ee
The one loop masses acquired by the pseudomoduli are
\bea
m_{Y Y^*}^2\! =\!  m_{\tilde Y \tilde Y^*}^2 =&&\!\!\!
\left((\cos\theta \cos \tau\!-\! \sin \theta \sin \tau)^2 ~(m^2\!+\!X^2)
+(\cos \theta  \sin \theta \!+\! \cos \theta \sin \tau)^2 \mu^2\right)~
\text{K}(m_A,m_2)\non\\
+&&\!
(\cos \theta \cos \tau - \sin \theta \sin \tau)^2~
\text{J}(m_A)
+
(\begin{array}{c}
\cos \theta \!\rightarrow \sin \theta,~
\sin \theta \!\rightarrow \!-\!\cos \theta,~
m_A \!\rightarrow m_B\end{array})
\non\\
+&&\! ((m_A,m_B) \rightarrow (m_C,m_D), ~ \theta \rightarrow \gamma)
\non\\
-&&\!
((\sin^2\tau (m^2+X^2)+\mu^2\cos^2 \tau)
\text{K}(m_F^{2-},m_2)+
(\begin{array}{cc}
\cos \tau \rightarrow \sin \tau,&
m_F^{2-} \rightarrow m_F^{2+}
\end{array}
)
)
\non\\
\eea
and
\bea
m_{Y\tilde Y}=
&&
2 m X
(\cos \theta \cos \tau - \sin \theta \sin \tau)^2 ~K(m_A,m_2)+
(\cos \theta \rightarrow \sin \theta, \sin \theta \rightarrow -\cos \theta,
\rightarrow m_A\rightarrow m_B)
\non \\
+&&((m_A,m_B)\rightarrow (m_C,m_D),\theta \rightarrow \gamma)
-\sin^2\tau K(m_F^{2-},m_2)-\cos^2\tau K(m_F^{2+},m_2))
\eea
where the functions K$(m_{B1},m_{B2})$ and J$(m_B)$ are associated with the Feynman diagrams of figure
\ref{massascala1} and \ref{massascala2}. The computation of these diagrams gives 
\bea
\text{K}(m_{B1},m_{B2})&&=-\frac{1}{16\pi^2}\left(
\log\frac{\Lambda^2}{m_{B2}^2}-\frac{m_{B1}^2}{m_{B1}^2-m_{B2}^2}
\log\frac{m_{B1}^2}{m_{B2}^2}
\right)\non \\ 
\text{J}(m_B) &&=\phantom{-} \frac{1}{16\pi^2}
\left(\Lambda^2-m_B^2 \log{\frac{\Lambda^2}{m_B^2}}\right)
\eea
In the $R$ symmetric limit, $m_{1}\rightarrow 0$, the vev of the 
scalar pseudomodulus $X$ vanishes. 
In this limit
the masses are
\bea
&&m_{Y Y^*}^2 =  m_{\tilde Y \tilde Y^*}^2
=
\\
&&=
\frac{
\mu^2(
(1-\nu^2)
(
(1-\epsilon^2)(\nu^2-1-\epsilon)\log(1-\epsilon)
+
(1+\epsilon^2)(\nu^2+\epsilon-1)\log(1+\epsilon)
)
-
4\epsilon^2\nu^4\log{\nu})
}
{32 \pi^2(\nu^2-1)(\epsilon^2-(\nu^2-1)^2)}\non
\eea
where we defined 
$\epsilon=\frac{f}{\mu^2}$ and $\nu=\frac{m}{\mu}$.
At the lowest order in the supersymmetry breaking scale this mass reduces to
\be
m_{Y Y^*}^2 =  m_{\tilde Y \tilde Y^*}^2=
\frac{f^2}{32 \pi^2 \mu^2}M(\nu)
=
\frac{f^2}{32 \pi^2 \mu^2} \frac{ (1- 4 \nu^2 + 3 \nu^4 - 4 \nu^4
  \log\nu) }{ (1-\nu^2)^3 }
\ee
where $M(\nu)$ is a positive function.

\subsubsection*{The flavored deformed $L^{131}$ model}

In section \ref{stringemb} we studied a different dark matter sector
in addiction to the superpotential (\ref{simplest}).
The superpotential for this DM sector is
\be
W_3=Y \phi_1 \phi_6 + \tilde Y \phi_2 \phi_5 
+ \mu_2 (\phi_5\phi_8+\phi_6\phi_7)+m_2\phi_7\phi_8
\ee
In this case the squared mass matrices of the fields
$\phi_5,\dots,\phi_8$ have to be rotated in
a diagonal form. This is done by defining the function
\be
\sin^2\alpha=\frac{\mu^2-\lambda_{-}^2}{\lambda_{+}^2-\lambda_{-}^2}
\ee
where 
\be
\lambda_{\pm}=\frac{
m_2^2 + 2\mu^2 - m \sqrt{m^2 + 4\mu^2}}{2}
\ee
The diagonal combinations $\rho_1,\dots,\rho_8$
appearing in $W_3$ are defined by
\bea
&&\phi_5=-\sin\alpha \rho_5+\cos\alpha \rho_8,\quad
\phi_8=\sin\alpha \rho_5+\cos\alpha \rho_8\non \\
&&\phi_6=-\sin\alpha \rho_6+\cos\alpha \rho_7,\quad
\phi_7=\cos\alpha \rho_5+\sin\alpha \rho_8
\eea
The fermion mass in this case is
\bea
M_{\psi_Y \psi_{\tilde Y}} =&& 
(\cos\theta \cos \tau - \sin \theta \sin \tau)^2 (\cos^2\alpha 
\ \text{I}(m_A,m_{f-})+ \sin^2\alpha \ \text{I}(m_A,m_{f+}))\non\\
+&&(\sin\theta \cos \tau + \cos \theta \sin \tau)^2  (\cos^2\alpha 
\ \text{I}(m_B,m_{f-})+ \sin^2\alpha \ \text{I}(m_B,m_{f+}))
\non \\ 
-&&(\cos\gamma \cos \tau + \sin \gamma \sin \tau)^2 (\cos^2\alpha 
\ \text{I}(m_C,m_{f-})+ \sin^2\alpha \ \text{I}(m_C,m_{f+}))
\non \\
-&&(\sin\gamma \cos \tau - \cos \gamma \sin \tau)^2  (\cos^2\alpha 
\ \text{I}(m_D,m_{f-})+ \sin^2\alpha \ \text{I}(m_D,m_{f+}))
\non\\\eea
where \be m_{f \pm} = \frac{m_2\pm\sqrt{m_2^2+4 \mu_2^2}}{2}\ee

\end{document}